\documentclass[%twocolumn,%showpacs,
preprintnumbers]{revtex4}
\usepackage{amsmath}
\usepackage{amsmath,amssymb}
\usepackage{bm}
\usepackage[dvips]{graphicx}
\begin{document}
%repeat the \author\address pair as needed
\preprint{\vtop{
{\hbox{YITP-18-65}\vskip-0pt
%                 \hbox{KANAZAWA-08-05} \vskip-0pt
%                 \hbox{hep-ph/??????} 
}}}
%}
%\preprint{\vtop{{\hbox{YITP-08-43}\vskip-0pt
%\hbox{hep-ph/08?????}
%}}}
%\date{\today}
\title{Open- and hidden-charm tetra-quark scalar mesons}
\author{Kunihiko Terasaki}
%\email[e-mail: ]{terasaki@hep.s.kanazawa-u.ac.jp,
%terasaki@yukawa.kyoto-u.ac.jp}
\affiliation{
Yukawa Institute for Theoretical Physics, Kyoto University, Kyoto 
606-8502, Japan
}
%\numberwithin{equation}{section}

\begin{abstract}
Masses of open- and hidden-charm tetra-quark scalar mesons are studied in a new 
approach. 
As the result, newly estimated masses of hidden-charm sector are significantly lower than 
our earier ones by a naive quark counting. 
In particular, our new result on the non-strange iso-triplet scalar mesons should be 
compared with the experiment that an indication of tiny $\eta\pi$ peak around 3.2 GeV 
has been observed, while no signal in the $\eta_c\pi$ channel. 
\end{abstract}
\maketitle

The charm-strange scalar $D_{s0}^+(2317)$ meson was discovered in the $D_s^+\pi^0$ 
mass distribution~\cite{D_{s0}(2317)-exp}, and its multi-quark interpretations have 
attracted general interests~\cite{Swanson}. 
For example, it was assigned to a tetra-quark meson, and its open- and 
hidden-charm partners were expected to exist~\cite{Cheng-Hou,Terasaki-D_s}. 
On the other hand, an indication of tiny peak around 3.2 GeV in the $\eta\pi$ invariant 
mass distribution was observed in two photon collisions~\cite{Uehara-delta(3.2)}. 
If it is a true signal, it can be a hidden-charm partner of $D_{s0}^+(2317)$, which was called 
as $\hat{\delta}^c(3200)$ in \cite{KT-delta(3.2)}. 
However, no resonance peak around the same energy has been observed in the 
$\eta_c\pi$ channel~\cite{Uehara-eta_c-pi}. 
It is strange, because the $\hat{\delta}^c(3200)\rightarrow \eta_c\pi$ decay is allowed 
under the OZI-rule~\cite{OZI} and the mass of $\hat{\delta}^c(3200)$ is sufficiently 
higher than the threshold, when its mass is truly 3200 MeV as 
measured~\cite{Uehara-delta(3.2)} (or $\sim 3300$ MeV as predicted by using a naive 
quark counting~\cite{KT-delta(3.2)}). 
This seems to suggest that its true mass is lower than the $\eta_c\pi$ threshold. 
In this short note, therefore, we study mass spectra of open- and hidden-charm 
tetra-quark scalar mesons from a new approach which is much different from our naive 
quark counting and discuss why partners of $D_{s0}^+(2317)$ (except for 
$\hat{\delta}^c(3200)$) have not been observed, and, in particular, why 
$\hat{\delta}^c(3200)$ was indicated in the $\eta\pi$ channel but not in the $\eta_c\pi$. 

Before going to mass spectra of open- and hidden-charm scalar mesons, we review very 
briefly on our picture of tetra-quark mesons. 
They can be classified into the following four groups~\cite{Jaffe}, 
%%%%%%%%%%%%%%%%%%%%%%%%%%%%%%%%%%%%%%%%%%%%%%%%%%%%%%%%%%%%%%%%%%%%%%%%%%
\begin{equation}
\{qq\bar q\bar q\} = [qq][\bar q\bar q] \oplus (qq)(\bar q\bar q) 
                            \oplus \{[qq](\bar q\bar q)\pm (qq)[\bar q\bar q]\},
                                                                                                \label{eq:4-quark}
\end{equation}
%%%%%%%%%%%%%%%%%%%%%%%%%%%%%%%%%%%%%%%%%%%%%%%%%%%%%%%%%%%%%%%%%%%%%%%%%%
according to different symmetry properties of their flavor wavefunctions, where 
parentheses and square brackets denote symmetry and anti-symmetry, respectively, 
under exchange of flavors between them. 
Here it should be noted that there are two ways to get a color-singlet tetra-quark 
$\{qq\bar{q}\bar{q}\}$ state, i.e., $\{qq\}$ (and $\{\bar{q}\bar{q}\}$) are (i) of 
$\bf{\bar 3_c}$ (and $\bf{3_c}$), and (ii) of $\bf{6_c}$ (and $\bf{\bar 6_c}$), respectively, 
of the color $SU_c(3)$. 
Therefore, each term in the right-hand side (r.h.s.) of Eq.~(\ref{eq:4-quark}) is again 
classified into two groups, i.e., $\{qq\}_{\bar{3}_c}\{\bar{q}\bar{q}\}_{3_c}$ and 
$\{qq\}_{6_c}\{\bar{q}\bar{q}\}_{\bar{6}_c}$. 
We here consider that the former is lower than the latter, because forces between two 
quarks (and between two antiquarks) are attractive when they are of $\bf{\bar 3_c}$ (and 
of $\bf{3_c}$), while repulsive when of $\bf{6_c}$ (and of $\bf{\bar 6_c}$)~\cite{Hori}. 
However, these $\{qq\}_{\bar{3}_c}\{\bar{q}\bar{q}\}_{3_c}$ and 
$\{qq\}_{6_c}\{\bar{q}\bar{q}\}_{\bar{6}_c}$ states can significantly mix with each other in 
the light sector ($q = u,\,d,\,s)$~\cite{Jaffe}.  
In contrast, such a mixing will be much smaller in heavy mesons, because the energy scale 
under consideration is much higher than the QCD scale ($\Lambda_{\rm QCD}$). 

Spin wave functions of $[qq]_{\bar{3}_c}$ and $[qq]_{{6}_c}$ should be anti-symmetric and 
symmetric in the flavor symmetry limit, because their wavefunctions should be totally 
anti-symmetric in the limit, i.e., they are of singlet ($\bm{1}_s$) and triplet ($\bm{3}_s$), 
respectively, of spin $SU(2)$. 
With regard to flavor symmetry breaking in meson wave functions, it is considered as 
follows. 
A matrix element of flavor charge which generates a flavor transformation is given by a  
form factor $f_+(0)$ in a relevant matrix element of vector current at zero momentum 
transfer squared, and therefore, a measure of the flavor symmetry breaking is given by 
a deviation of its value from unity~\cite{TM}. 
By the way, values of various $f_+(0)$'s have been summarized in \cite{PDG96}. 
As the first example, the chiral perturbation theory has provided 
%%%%%%%%%%%%%%%%%%%%%%%%%%%%%%%%%%%%%%%%%%%%%%%%%%%%%%%%%%%%%%%%%%%%%%%%%%
${f_+^{(\pi K)} (0)\, =  0.961 \pm 0.008}$~\cite{Leutwyler}  
%%%%%%%%%%%%%%%%%%%%%%%%%%%%%%%%%%%%%%%%%%%%%%%%%%%%%%%%%%%%%%%%%%%%%%%%%%
which implies that the flavor $SU_f(3)$ symmetry works well in meson wave functions. 
In addition, it works well even in the world of charm mesons, as seen in 
%%%%%%%%%%%%%%%%%%%%%%%%%%%%%%%%%%%%%%%%%%%%%%%%%%%%%%%%%%%%%%%%%%%%%%%%%%
${f_+^{(\pi D)}(0)/f_+^{(\bar{K}D)}(0)} = {1.00 \pm 0.11 \pm 0.02}$~\cite{FNAL-E687} and 
${0.99 \pm 0.08}$~\cite{CLEO-FF}.  
%%%%%%%%%%%%%%%%%%%%%%%%%%%%%%%%%%%%%%%%%%%%%%%%%%%%%%%%%%%%%%%%%%%%%%%%%%
Although the measured $f_+^{(\bar{K} D)}(0) =  0.74 \pm 0.03$~\cite{PDG96} suggests that 
the $SU_f(4)$ symmetry in wave functions is broken, its deviation from unity is not fatally 
large, and therefore, we take wave functions of open- and hidden-charm tetra-quark 
mesons in the flavor $SU_f(4)$ symmetry limit (as their dominant part) in this short note. 
In this case, the first term $[qq][\bar q\bar q]$ in the r.h.s. of Eq.~(\ref{eq:4-quark}) has 
(dominantly) $J^P = 0^+$, where $J$ and $P$ denote its spin and parity, respectively. 

Now it is known that the mass spectrum of the observed light scalar nonoet, 
$f_0(500)$, $a_0(980)$, $f_0(980)$~\cite{PDG06} and $\kappa(800)$~\cite{E791} is well 
reproduced by the light tetra-quark $[qq][\bar{q}\bar{q}]$ states~\cite{Jaffe}, and it is 
easy to extend this framework to open- and hidden-charm scalar mesons, 
$[cq][\bar{q}\bar{q}]$ and $[cq][\bar{c}\bar{q}]$, putting their detailed structure of color 
wavefunctions aside. 
In fact, just after the discovery of $D_{s0}^+(2317)$, it was assigned to the iso-singlet 
$\{[cn][\bar s\bar n]\}_{I=0},\,(n=u,d)$ state~\cite{Cheng-Hou} to understand its narrow 
width. 
If $D_{s0}^+(2317) (\sim \{[cn][\bar s\bar n]\}_{I=0}) \rightarrow D_s^+\pi^0$ through the 
isospin nonconserving hadronic interactions were dominant in its decays, its rate would be 
small, because such interactions are much weaker than the isospin conserving strong 
ones, and therefore, its narrow width could be understood. 
In this case, however, its dominant decay should be the radiatve 
$D_{s0}^+(2317)\rightarrow D_s^{*+}\gamma$~\cite{HT-isospin}, because of the hierarchy 
of hadron interactions~\cite{Dalitz},  
%%%%%%%%%%%%%%%%%%%%%%%%%%%%%%%%%%%%%%%%%%%%%%%%%%%%%%%%%%%%%%%%%%%%%%%%%%
\begin{equation}
|{\rm isospin\,\, conserving\,\, int.} \sim O(1)| \gg |{\rm electromagnetic\,\, int.} 
\sim O(\sqrt{\alpha})| \gg  
|{\rm isospin\,\, nonconserving\,\, hadronic\,\, int.} \sim O(\alpha)|, \nonumber
\end{equation}
%%%%%%%%%%%%%%%%%%%%%%%%%%%%%%%%%%%%%%%%%%%%%%%%%%%%%%%%%%%%%%%%%%%%%%%%%%
unless there exists any specific mechanism to suppress strongly the $D_s^{*+}\gamma$ 
decay and/or to enhance extremely the $D_s^+\pi^0$, where $\alpha$ is the fine 
structure constant. 
However, such a mechanism is not known in any tetra-quark picture of $D_{s0}^+(2317)$ 
yet. 
Therefore, it is strange that no signal of $D_{s0}^+(2317)$ has been observed in the 
$D_s^{*+}\gamma$ channel. 
In addition, it is expected that its iso-triplet partners which decay dominantly into 
$D_s^+\pi$ final states exist. 
In this case, experiments should have observed two peaks with approximately degenerate 
masses but distinctly different widths in the $D_s^+\pi^0$ channel, unless production of 
the iso-triplet compornent is strongly suppressed. 
Nevertheless, such states have not been observed yet. 
In contrast, the present author assigned it to the iso-triplet 
%%%%%%%%%%%%%%%%%%%%%%%%%%%%%%%%%%%%%%%%%%%%%%%%%%%%%%%%%%%%%%%%%%%%%%%%%%
$\{[cn]_{\bar{3}_c}^{1_s}[\bar s\bar n]_{3_c}^{1_s}\}_{1_c}^{1_s}|_{I=1}$ 
%%%%%%%%%%%%%%%%%%%%%%%%%%%%%%%%%%%%%%%%%%%%%%%%%%%%%%%%%%%%%%%%%%%%%%%%%%
state~\cite{Terasaki-D_s}. 
In this case, its dominant decay is expected to be the isospin conserving 
$D_{s0}^+(2317)\rightarrow D_s^+\pi^0$. 
Although it is a strong decay, its rate can be small, because of a small overlap of 
wavefunctions between the initial 
%%%%%%%%%%%%%%%%%%%%%%%%%%%%%%%%%%%%%%%%%%%%%%%%%%%%%%%%%%%%%%%%%%%%%%%%%%
$D_{s0}^+(2317)\sim \{[cn]_{\bar{3}_c}^{1_s}[\bar s\bar n]_{3_c}^{1_s}\}_{1_c}^{1_s}|_{I=1}$ 
%%%%%%%%%%%%%%%%%%%%%%%%%%%%%%%%%%%%%%%%%%%%%%%%%%%%%%%%%%%%%%%%%%%%%%%%%%
and the final $D_s^+\pi^0$, as was discussed in \cite{HT-isospin} and will be seen later 
again.  
It seems to be nice, while no signal of its neutral and doubly charged partners, 
$D_{s0}^0(2317)$ and $D_{s0}^{++}(2317)$, has been observed in $B$ 
decays~\cite{search-for-double-charge}. 
It might imply that their production rates are much lower than that for $D_{s0}^+(2317)$, 
against our expectation~\cite{KT-production}. 

Putting this problem aside for a while, we now study masses of open- and hidden-charm 
scalar tetra-quark mesons. 
Before going to the main subject, we study masses of ordinary mesons. 
An ideally mixed ordinary meson which is considered as a color singlet quark-antiquark 
$\{q_i\bar{q}_j\}_{1_c}$ system has a mass which is given by a sum of spin averaged mass 
$M(\{q_i\bar{q}_j\}_{1_c})$ and hyperfine splitting of the $\{q_i\bar{q}_j\}_{1_c}$ 
pair~\cite{KNR},  
%%%%%%%%%%%%%%%%%%%%%%%%%%%%%%%%%%%%%%%%%%%%%%%%%%%%%%%%%%%%%%%%%%%%%%%%%%
\begin{equation}
M(\{{\rm meson}\}_{ij}) = 
M(\{q_i\bar{q}_j\}_{1_c}) + a_{ij}\frac{\bm{\sigma}_i \cdot \bm{\sigma}_j}{m_im_j}, 
                                                                                  \label{eq:meson-mass-gen}
\end{equation}
%%%%%%%%%%%%%%%%%%%%%%%%%%%%%%%%%%%%%%%%%%%%%%%%%%%%%%%%%%%%%%%%%%%%%%%%%%
where $m_i$ is the mass of quark $q_i = u,\,d,\,s,\,{\rm or}\,\,c$ and its spin is given by 
$\frac{1}{2}\bm{\sigma}_i$. 
In this case, masses of  ideally mixed mesons are given by  
%%%%%%%%%%%%%%%%%%%%%%%%%%%%%%%%%%%%%%%%%%%%%%%%%%%%%%%%%%%%%%%%%%%%%%%%%%
\begin{eqnarray}
&&\hspace{-5mm}
\left\{\,\,\begin{tabular}{l}
\quad\,\, $M(J/\psi)\quad = \,M(\{c\bar{c}\}_{1_c})\, +\,\,\, 2a_{cc}/(m_c)^2$\,\,\,, 
\quad $M(\eta_{c\,})\,\, = \, M(\{c\bar{c}\}_{1_c})\, - \,\,\,\,6a_{cc}/(m_c)^2$\,\,, \\
\quad\,\, $M(D_{s\,}^{*\,})\quad\,\, = \, M(\{c\bar{s}\}_{1_c})\, +\, 2a_{cs}/(m_cm_s)$,  
\quad $M(D_{s}) =  \,M(\{c\bar{s}\}_{\,1_c}) - \,6a_{cs}/(m_cm_s)$, \\
\quad\,\, $M(D^{\,*\,})\quad\, =   M(\{c\bar{n}\}_{1_c})\, + 2a_{cn}/(m_cm_n)$, 
\quad$M(D)\,\, =  \,M(\{c\bar{n}\}_{1_c}) - 6a_{cn}/(m_cm_n)$, \\
\quad\,\, $M(\bar{K}^{*})\quad\,\, = M(\{n\bar{s}\}_{1_c\,}) + 2a_{sn}/(m_sm_n)$,  
\quad $M(\bar{K})\, = \,M(\{n\bar{s}\}_{1_c}) - 6a_{sn}/(m_sm_n)$, \\
$M(\rho) = M(\omega) = M(\{n\bar{n}\}_{1_c}) + \,\,2a_{nn}/(m_n)^2$\,\,, 
\quad\,$M(\pi\,)\, =  \,M(\{n\bar{n}\}_{1_c}) - \,\,\, 6a_{nn}/(m_n)^2$\,, 
\end{tabular}\right.
                                                                         \label{eq:low-lying-meson-mass}
\end{eqnarray}
%%%%%%%%%%%%%%%%%%%%%%%%%%%%%%%%%%%%%%%%%%%%%%%%%%%%%%%%%%%%%%%%%%%%%%%%%%
where $\{s\bar{s}\}$ mesons have been excluded in the above, because the $\eta\eta'$ 
mixing is far from the ideal one.  
In this way, the spin averaged masses %and the hyperfine splittings 
are determined as 
%%%%%%%%%%%%%%%%%%%%%%%%%%%%%%%%%%%%%%%%%%%%%%%%%%%%%%%%%%%%%%%%%%%%%%%%%%
\begin{eqnarray}
&&\hspace{-8mm}
\left\{\begin{tabular}{l}
\,$\bar{M}(\{c\bar{c}\}_{1_c})\, = [3M(J/\psi) + M(\eta_c)]/4 = 3068.5$ MeV,\quad 
\,$\bar{M}(\{c\bar{s}\}_{1_c})\, = [3M(D_s^*) + M(D_s)]/4 \,= 2076.4$ MeV, \\
\,$\bar{M}(\{c\bar{n}\}_{1_c}) = \,[3M(D^*) + M(D)]/4\,\, = 1973.3$ MeV, \,\,\, 
$\bar{M}(\{s\bar{n}\}_{1_c}) = \,[3M(\bar{K}^*) + M(\bar{K})]/4\,=\,\, \,794.5$ MeV, \\
$\bar{M}(\{n\bar{n}\}_{1_c}) = \{3[M(\omega) + M(\rho)]/2 + M(\pi)\}/4
\hspace{-0.3mm}=\,\,\, 618.4$ MeV, 
\end{tabular}\right.                                                         \label{eq:mass-of-q-anti-q)}
\end{eqnarray}
%%%%%%%%%%%%%%%%%%%%%%%%%%%%%%%%%%%%%%%%%%%%%%%%%%%%%%%%%%%%%%%%%%%%%%%%%%
by taking the measured mass values of mesons in \cite{PDG16}, where their 
small errors have been neglected.  

The spin averaged mass $M(\{q_i\bar{q}_j\}_{1_c})$ is given by a sum of masses of quarks 
which construct the system and their binding energy, and the so-called constituent quark 
masses have been taken as those of the constituents in \cite{KNR}. 
In this short note, however, their running masses are taken and, at the same time, a 
confinement effect (including possible gluon contributions) in the ordinary meson is 
parametrized by $G$, 
%%%%%%%%%%%%%%%%%%%%%%%%%%%%%%%%%%%%%%%%%%%%%%%%%%%%%%%%%%%%%%%%%%%%%%%%%%
\begin{equation}
M(\{q_i\bar{q}_j\}_{1_c})  = m_i + m_j + G + B(\{q_i\bar{q}_j\}_{1_c}),  
\end{equation}
%%%%%%%%%%%%%%%%%%%%%%%%%%%%%%%%%%%%%%%%%%%%%%%%%%%%%%%%%%%%%%%%%%%%%%%%%%
where $G$ is assumed to be flavor independent. 
To estimate the binding energy $B(\{q_i\bar{q}_j\}_{1_c})$, we take the following values of 
running quark masses; 
%%%%%%%%%%%%%%%%%%%%%%%%%%%%%%%%%%%%%%%%%%%%%%%%%%%%%%%%%%%%%%%%%%%%%%%%%%
$m_n(\mu = 2\,{\rm GeV}) = \bigl(3.5^{+0.7}_{-0.3}\bigr)$ MeV, %\\
$m_s(\mu = 2\,{\rm GeV}) %\hspace{0mm}
= \bigl(96^{+8}_{-4}\bigr)$ MeV, %\\
$m_c(\mu = m_c) = (1280\pm 30)$ MeV~\cite{PDG16} and  %\\                
$m_{c}(\mu = 3\,{\rm GeV}) = (993\pm 8)$ MeV~\cite{m_c(3GeV)}.   
%%%%%%%%%%%%%%%%%%%%%%%%%%%%%%%%%%%%%%%%%%%%%%%%%%%%%%%%%%%%%%%%%%%%%%%%%%
We here assume, for simplicity, that $m_n(\mu)$ and $m_s(\mu)$ are rather stable (do 
not so rapidly run) in the energy scale of open- and hidden-charm meson masses under 
consideration, because their contribution is not very large and, as the result, this 
assumption does not make serious errors in the present work. 
In addition, it is assumed that $m_c(\mu \sim 2\,\,{\rm GeV})$ is not very much 
different from $m_c(\mu = m_c)$. 
By taking the above values of quark masses, spin averaged binding energies 
$B(\{q\bar{q}\}_{1_c})$'s of color singlet $\{q\bar{q}\}_{1_c}$ states at the energy scale 
($\mu \sim 2$ or $3$ GeV) under consideration can be crudely obtained as 
%%%%%%%%%%%%%%%%%%%%%%%%%%%%%%%%%%%%%%%%%%%%%%%%%%%%%%%%%%%%%%%%%%%%%%%%%%
\begin{eqnarray}
&&\hspace{-3mm}
\left\{\begin{tabular}{l}
\,\,\,$B(\{c\bar{c}\}_{1_c})\, =\,\,  1082.5\qquad\,\,\,
\,\,{\rm MeV}- G$,\quad     
\,\,\,\,$B(\{c\bar{s}\}_{\,1_c})\, = \,\,700.4\pm 30\,\,{\rm MeV} - G$, \\    
\,\,$B(\{c\bar{n}\}_{1_c})\, = \,\,\,\,\,689.8\pm 30\,\,\,{\rm MeV} - G$, \quad 
\,\,$B(\{n\bar{s}\}_{1_{c}})\, = \,\,695.0\qquad\,\,\,{\rm MeV} - G$,\\
\,$B(\{n\bar{n}\}_{1_c})\, = \,\,\,\,\,611.4\qquad\,\,\,\,{\rm MeV} - G$
\end{tabular}\right.                                                           \label{eq:binding-energy}
\end{eqnarray}
%%%%%%%%%%%%%%%%%%%%%%%%%%%%%%%%%%%%%%%%%%%%%%%%%%%%%%%%%%%%%%%%%%%%%%%%%%
under the isospin $SU_I(2)$ symmetry, where errors from $m_n(\mu = 2\,{\rm GeV})$ and 
$m_s(\mu = 2\,{\rm GeV})$ have been neglected, because they do not make serious 
uncertainties in the above results. 

Next, we study masses of open- and hidden-charm tetra-quark scalar mesons in a new 
picture similar to a QCD-string-junction model~\cite{KNR}. 
In this picture, a mass of tetra-quark 
$\{[q_iq_j]_{\bar{3}_c}^{1_s}[\bar{q}_k\bar{q}_\ell]_{3_c}^{1_s}\}_{1_c}^{1_s}$ 
meson is given in the form 
%%%%%%%%%%%%%%%%%%%%%%%%%%%%%%%%%%%%%%%%%%%%%%%%%%%%%%%%%%%%%%%%%%%%%%%%%%
\begin{eqnarray}
&&\hspace{-3mm}
M(\{[q_iq_j]_{\bar{3}_c}^{1_s}[\bar{q}_k\bar{q}_\ell]_{3_c}^{1_s}\}_{1_c}^{1_s}) 
= 2\hat{S} + (m_i + m_j + m_k + m_\ell) 
  + \hat{G} + B(\{[q_iq_j]_{\bar{3}_c}^{1_s}[\bar{q}_k\bar{q}_\ell]_{3_c}^{1_s}\}_{1_c}^{1_s}),
                                                                                      \label{eq:string-junction}
\end{eqnarray}
%%%%%%%%%%%%%%%%%%%%%%%%%%%%%%%%%%%%%%%%%%%%%%%%%%%%%%%%%%%%%%%%%%%%%%%%%%
where $\hat{S}$ and $\hat{G}$ denote a QCD-string-junction contribution and a 
confinement effect in the tetra-quark system. 
In the above equation, we have neglected a contribution of spin-spin force between 
the so-called diquark $[q_iq_j]_{\bar{3}_c}$ and antidiquark $[\bar{q}_k\bar{q}_\ell]_{3_c}$, 
because their spins are vanishing in the present case, in contrast to \cite{KNR}. 
To compute the binding energies 
$B(\{[qq]_{\bar{3}_c}^{1_s}[\bar{q}\bar{q}]_{3_c}^{1_s}\}_{1_c}^{1_s})$'s of 
tetra-quark scalar mesons, we first decompose each of tetra-quark states into a sum of 
products of $\{q\bar{q}\}$ pairs, as in \cite{KT-delta(3.2),KT-decomp}. 
(Our notations $\hat{F}_{I}$, $\hat{F}_{0}$, $\hat{D}$, ($\hat{D}^s$), $\hat{E}^0$, 
$\hat{\kappa}^c$, $\hat{\delta}^c$, ($\hat{\sigma}^{cs}$) and $\hat{\sigma}^c$ of the 
open- and hidden-charm tetra-quark scalar mesons have been provided in 
\cite{Terasaki-D_s} and \cite{KT-delta(3.2)}, respectively, where $\hat{D}^s$ and 
$\hat{\sigma}^{cs}$ are not considered in this short note, because each of them contains 
an $\{s\bar{s}\}$ pair.)  
To save space, we here list only the result on $\hat{F}_I^+$ which is 
assigned~\cite{Terasaki-D_s} to $D_{s0}^+(2317)$, 
%%%%%%%%%%%%%%%%%%%%%%%%%%%%%%%%%%%%%%%%%%%%%%%%%%%%%%%%%%%%%%%%%%%%%%%%%%
\begin{eqnarray}
&&\hspace{-5mm}
|\hat{F}_{I}^+ \rangle 
= \frac{1}{2}\sqrt{\frac{1}{2}}\bigl|
                          [cu]_{\bar{3}_c}^{1_s}[\bar{s}\bar{u}]_{3_c}^{1_s} 
                          -  [cd]_{\bar{3}_c}^{1_s}[\bar{s}\bar{d}]_{3_c}^{1_s}\bigr\rangle_{1_c}^{1_s} 
\nonumber\\
&&\hspace{-3mm}
= \frac{1}{2}\sqrt{\frac{1}{2}}\Bigl\{
- \sqrt{\frac{1}{12}}
                   \bigl|\{c\bar{s}\}_{1_c}^{1_s} \{u\bar{u}\} _{1_c}^{1_s}\bigl\rangle_{1_c}^{1_s} 
+ \sqrt{\frac{3}{12}}
                   \bigl|\{c\bar{s}\}_{1_c}^{3_s} \{u\bar{u}\} _{1_c}^{3_s}\bigl\rangle_{1_c}^{1_s}  
+ \sqrt{\frac{1}{12}}\bigl|\{c\bar{u}\}_{1_c}^{1_s} \{u\bar{s}\} _{1_c}^{1_s}
                                                                                     \bigl\rangle_{1_c}^{1_s} 
- \sqrt{\frac{3}{12}}\bigl|\{c\bar{u}\}_{1_c}^{3_s} \{u\bar{s}\} _{1_c}^{3_s}
                                                                                     \bigl\rangle_{1_c}^{1_s}  
\nonumber\\
&&\hspace{12mm}
+ \sqrt{\frac{1}{12}}\bigl|\{u\bar{s}\}_{1_c}^{1_s} \{c\bar{u}\} _{1_c}^{1_s}
                                                                                    \bigl\rangle_{1_c}^{1_s} 
- \sqrt{\frac{3}{12}}\bigl|\{u\bar{s}\}_{1_c}^{3_s} \{c\bar{u}\} _{1_c}^{3_s}
\bigl\rangle_{1_c}^{1_s} 
- \sqrt{\frac{1}{12}}\bigl|\{u\bar{u}\}_{1_c}^{1_s} \{c\bar{s}\} _{1_c}^{1_s}
\bigl\rangle_{1_c}^{1_s} 
+ \sqrt{\frac{3}{12}}\bigl|\{u\bar{u}\}_{1_c}^{3_s} \{c\bar{s}\} _{1_c}^{3_s}
\bigl\rangle_{1_c}^{1_s} 
\nonumber\\
&&\hspace{12mm}
+ \sqrt{\frac{1}{12}}\bigl|\{c\bar{s}\}_{1_c}^{1_s} \{d\bar{d}\} _{1_c}^{1_s}
\bigl\rangle_{1_c}^{1_s} 
- \sqrt{\frac{3}{12}}\bigl|\{c\bar{s}\}_{1_c}^{3_s} \{d\bar{d}\} _{1_c}^{3_s}
\bigl\rangle_{1_c}^{1_s} 
- \sqrt{\frac{1}{12}}\bigl|\{c\bar{d}\}_{1_c}^{1_s} \{d\bar{s}\} _{1_c}^{1_s}
\bigl\rangle_{1_c}^{1_s} 
+ \sqrt{\frac{3}{12}}\bigl|\{c\bar{d}\}_{1_c}^{3_s} \{d\bar{s}\} _{1_c}^{3_s}
\bigl\rangle_{1_c}^{1_s} 
\nonumber\\
&&\hspace{12mm}
- \sqrt{\frac{1}{12}}\bigl|\{d\bar{s}\}_{1_c}^{1_s} \{c\bar{d}\} _{1_c}^{1_s}
\bigl\rangle_{1_c}^{1_s} 
+ \sqrt{\frac{3}{12}}\bigl|\{d\bar{s}\}_{1_c}^{3_s} \{c\bar{d}\} _{1_c}^{3_s}
\bigl\rangle_{1_c}^{1_s} 
+ \sqrt{\frac{1}{12}}\bigl|\{d\bar{d}\}_{1_c}^{1_s} \{c\bar{s}\} _{1_c}^{1_s}
\bigl\rangle_{1_c}^{1_s} 
- \sqrt{\frac{3}{12}}\bigl|\{d\bar{d}\}_{1_c}^{3_s} \{c\bar{s}\} _{1_c}^{3_s}
\bigl\rangle_{1_c}^{1_s} 
\nonumber\\
&&\hspace{12mm}
- \sqrt{\frac{2}{12}}\bigl|\{c\bar{s}\}_{8_c}^{1_s}\{u\bar{u}\}_{8_c}^{1_s} 
\bigl\rangle_{1_c}^{1_s} 
+ \sqrt{\frac{6}{12}}\bigl|\{c\bar{s}\}_{8_c}^{3_s}\{u\bar{u}\}_{8_c}^{3_s} 
\bigl\rangle_{1_c}^{1_s} 
+ \sqrt{\frac{2}{12}}\bigl|\{c\bar{u}\}_{8_c}^{1_s}\{u\bar{s}\}_{8_c}^{1_s} 
\bigl\rangle_{1_c}^{1_s} 
- \sqrt{\frac{6}{12}}\bigl|\{c\bar{u}\}_{8_c}^{3_s}\{u\bar{s}\}_{8_c}^{3_s} 
\bigl\rangle_{1_c}^{1_s} 
\nonumber\\
&&\hspace{12mm}
+ \sqrt{\frac{2}{12}}\bigl|\{u\bar{s}\}_{8_c}^{1_s}\{c\bar{u}\}_{8_c}^{1_s} 
\bigl\rangle_{1_c}^{1_s} 
- \sqrt{\frac{6}{12}}\bigl|\{u\bar{s}\}_{8_c}^{3_s}\{c\bar{u}\}_{8_c}^{3_s} 
\bigl\rangle_{1_c}^{1_s} 
- \sqrt{\frac{2}{12}}\bigl|\{u\bar{u}\}_{8_c}^{1_s}\{c\bar{s}\}_{8_c}^{1_s} 
\bigl\rangle_{1_c}^{1_s} 
+ \sqrt{\frac{6}{12}}\bigl|\{u\bar{u}\}_{8_c}^{3_s}\{c\bar{s}\}_{8_c}^{3_s} 
\bigl\rangle_{1_c}^{1_s} 
\nonumber\\
&&\hspace{12mm}
+ \sqrt{\frac{2}{12}}\bigl|\{c\bar{s}\}_{8_c}^{1_s}\{d\bar{d}\}_{8_c}^{1_s} 
\bigl\rangle_{1_c}^{1_s} 
-  \sqrt{\frac{6}{12}}\bigl|\{c\bar{s}\}_{8_c}^{3_s}\{d\bar{d}\}_{8_c}^{3_s} 
\bigl\rangle_{1_c}^{1_s} 
- \sqrt{\frac{2}{12}}\bigl|\{c\bar{d}\}_{8_c}^{1_s}\{d\bar{s}\}_{8_c}^{1_s} 
\bigl\rangle_{1_c}^{1_s} 
+ \sqrt{\frac{6}{12}}\bigl|\{c\bar{d}\}_{8_c}^{3_s}\{d\bar{s}\}_{8_c}^{3_s} 
\bigl\rangle_{1_c}^{1_s} 
\nonumber\\
&&\hspace{12mm}
- \sqrt{\frac{2}{12}}\bigl|\{d\bar{s}\}_{8_c}^{1_s} \{c\bar{d}\}_{8_c}^{1_s}
\bigl\rangle_{1_c}^{1_s} 
+ \sqrt{\frac{6}{12}}\bigl|\{d\bar{s}\}_{8_c}^{3_s} \{c\bar{d}\}_{8_c}^{3_s}
\bigl\rangle_{1_c}^{1_s} 
+ \sqrt{\frac{2}{12}}\bigl|\{d\bar{d}\}_{8_c}^{1_s}\{c\bar{s}\}_{8_c}^{1_s} 
\bigl\rangle_{1_c}^{1_s} 
- \sqrt{\frac{6}{12}}\bigl|\{d\bar{d}\}_{8_c}^{3_s}\{c\bar{s}\}_{8_c}^{3_s} 
\bigl\rangle_{1_c}^{1_s} 
\Bigr\},  
\nonumber\\
&&\hspace{-3mm}
= \frac{1}{4}\sqrt{\frac{1}{6}}\Bigl\{
- \sqrt{2}\bigl|D_s^+\pi^0 + \pi^0D_s^+\bigr\rangle 
+ \sqrt{6}\bigl|D_s^{*+}\rho^0 + \rho^0D_s^{*+}\bigr\rangle 
+ \bigl|D^0K^+ + K^+D^0\bigr\rangle 
\nonumber\\
&&\hspace{12mm}
- \bigl|D^+K^0 + K^0D^+\bigr\rangle - \sqrt{3}\bigl|K^{*+}D^{*0} + D^{*0}K^{*+}\bigr\rangle 
+ \sqrt{3}\bigl|D^{*+}K^{*0} + K^{*0}D^{*+}\bigr\rangle \Bigr\} + \cdots,              
                                                                                           \label{eq:decomp-F_I}
\end{eqnarray}
%%%%%%%%%%%%%%%%%%%%%%%%%%%%%%%%%%%%%%%%%%%%%%%%%%%%%%%%%%%%%%%%%%%%%%%%%%
where $\{u\bar{s}\}_{1_c}^{1_s}$, $\{c\bar{s}\}_{1_c}^{1_s}$, $\{u\bar{s}\}_{1_c}^{3_s}$, 
$\{c\bar{s}\}_{1_c}^{3_s}$, etc. have been replaced by $K^+$, $D_s^+$, $K^{*+}$, 
$D_s^{*+}$, etc., respectively, and the ellipsis denotes neglected contributions of 
products of color octet $\{q\bar{q}\}_{8_c}$ pairs in the last equality. 
We here assume that the above decomposition is stable (not easily reshuffled by 
exchanging gluons), because the energy scale under consideration is much higher than 
$\Lambda_{QCD}$. 
In this way, it is seen that a wavefunction overlap between the initial 
$D_{s0}^+(2317)(= \hat{F}_I^+)$ and the final $D_s^+\pi^0$ is small, so that a rate for the 
main decay $D_{s0}^+(2317) \rightarrow D_s^+\pi^0$ can be small, and as the result, 
$D_{s0}^+(2317)$ can be narrow, as discussed before. 
In addition, square of coefficient of each $|\{q\bar{q}\}\{q\bar{q}\}\rangle$ state provides 
its share in the tetra-quark state under consideration. 
By noting the above decomposition of $\hat{F}_I$ (and those of its partners) and by taking 
the binding energies $B(\{q_i\bar{q}_j\}_{1_c})$'s in Eq.~(\ref{eq:binding-energy}), 
$B(\{q_i\bar{q}_j\}_{8_c})$'s which are equated to $-\frac{1}{8}B(\{q_i\bar{q}_j\}_{1_c})$'s 
(because the force between $q$ and $\bar{q}$ with the color $\bm{8}_c$ is $-\frac{1}{8}$ 
of that with $\bm{1}_c$~\cite{Hori}) and the binding energy $B_{8_c8_c}$ between two 
$\{q_i\bar{q}_j\}_{8_c}$ pairs, the total binding energies 
$B(\{[qq]_{\bar{3}_c}^{1_s}[\bar{q}\bar{q}]_{3_c}^{1_s}\rangle_{1_c}^{1_s})$'s can be obtained as 
%%%%%%%%%%%%%%%%%%%%%%%%%%%%%%%%%%%%%%%%%%%%%%%%%%%%%%%%%%%%%%%%%%%%%%%%%%
\begin{eqnarray}
&&\hspace{-10mm}
\left\{\begin{tabular}{l}
$\displaystyle{B(\hat{F}_{I}^+) 
= (337.1\pm 5.3)\,\,{\rm MeV} - \frac{G}{2} + \frac{2}{3}B_{8_c8_c}}$,\quad 
$\displaystyle{B(\hat{F}_{0}^+) 
= (337.1\pm 5.3)\,\,{\rm MeV} - \frac{G}{2} + \frac{2}{3}B_{8_c8_c}}$, \vspace{1mm}\\
\,\,$\displaystyle{B(\hat{D})\, 
= (325.3\pm 7.5)\,\,{\rm MeV} - \frac{G}{2} + \frac{2}{3}B_{8_c8_c}}$,\quad 
\,$\displaystyle{B(\hat{E}^{0\,}) 
= (346.2\pm 7.5)\,\,{\rm MeV} - \frac{G}{2} + \frac{2}{3}B_{8_c8_c}}$, \vspace{1mm}\\
\,$\displaystyle{B(\hat{\delta}^{c\,})\,
= (384.2\pm 7.5)\,\,{\rm MeV} - \frac{G}{2} + \frac{2}{3}B_{8_c8_c}}$,\quad 
\,$\displaystyle{B(\hat{\sigma}^{c\,}) \,
= (384.2\pm 7.5)\,\,{\rm MeV} - \frac{G}{2} + \frac{2}{3}B_{8_c8_c}}$, \vspace{1mm}\\
\,$\displaystyle{B(\hat{\kappa}^c)\, 
= (396.0\pm 5.3)\,\,{\rm MeV} - \frac{G}{2} + \frac{2}{3}B_{8_c8_c}}$.  
\end{tabular}\right.                                                    \label{eq:total-binding-energies}
\end{eqnarray}
%%%%%%%%%%%%%%%%%%%%%%%%%%%%%%%%%%%%%%%%%%%%%%%%%%%%%%%%%%%%%%%%%%%%%%%%%%
Insertion of Eq.~(\ref{eq:total-binding-energies}) into Eq.~(\ref{eq:string-junction}) leads 
to the following masses of open- and hidden-charm tetra-quark scalars,  
%%%%%%%%%%%%%%%%%%%%%%%%%%%%%%%%%%%%%%%%%%%%%%%%%%%%%%%%%%%%%%%%%%%%%%%%%%
\begin{eqnarray}
&&\hspace{-10mm}
\left\{\begin{tabular}{l}
$\,M(\hat{F}_{I}) = 2\hat{S} + (\hat{G} + m_c + m_s + 2m_n) + B(\hat{F}_I) 
= M(\hat{F}_0^+)$ \\  \hspace{10.5mm}
$= (1720.1\pm 30.5)\,\,{\rm MeV} 
                         + [2\hat{S} + \hat{G} - \frac{1}{2}G + \frac{2}{3}B_{8_c8_c}]$, \\
\,$M(\hat{D\,}) = 2\hat{S} + (\hat{G} + m_c + 3m_n) + B(\hat{D})$ \\ \hspace{10.5mm}
$= (1615.8\pm 30.9) \,\,{\rm MeV} 
                      + [2\hat{S} + \hat{G} - \frac{1}{2}G + \frac{2}{3}B_{8_c8_c}]$, \\
$M(\hat{E}^{0})\hspace{-0.2mm}
= 2\hat{S} + (\hat{G} + m_c + m_s + 2m_n) + B(\hat{E}^0)$ \\   
\hspace{10.5mm}
$= (1729.2\pm 30.9)\,\,{\rm MeV} 
                        + [2\hat{S} + \hat{G} - \frac{1}{2}G + \frac{2}{3}B_{8_c8_c}]$, \\
\,$M(\,\hat{\delta}^{c}) 
                   = 2\hat{S} + (\hat{G} + 2m_c + 2m_n) + B(\hat{\delta}^c) 
                   = M(\hat{\sigma}^c)$ \\
\hspace{11.5mm}$= (2383.2\pm\,\, 7.5)\,\,{\rm MeV}  
                        + [2\hat{S} + \hat{G} - \frac{1}{2}G + \frac{2}{3}B_{8_c8_c}]$, \\ 
$M(\hat{\kappa}^{c})\, 
= 2\hat{S} + (\hat{G} + 2m_c + m_s + m_n) + B(\hat{\kappa}^c)$\\
\hspace{10.3mm}
$= (2481.5\pm \,\,5.3)\,\,{\rm MeV}   
                        + [2\hat{S} + \hat{G} - \frac{1}{2}G + \frac{2}{3}B_{8_c8_c}]$.  
\end{tabular}\right.                                                            \label{eq:mass-formulae}
\end{eqnarray}
%%%%%%%%%%%%%%%%%%%%%%%%%%%%%%%%%%%%%%%%%%%%%%%%%%%%%%%%%%%%%%%%%%%%%%%%%%
Because of $B(\hat{F}_{I}) = B(\hat{F}_{0}^+)$, masses of $\hat{F}_I$ and $\hat{F}_0$ are 
degenerate in the present approach, as in our earier work~\cite{Terasaki-D_s}. 
The above masses of open- and hidden-charm tetra-quark scalar mesons include  
a combination of unknown paramters, 
$2\hat{S} + \hat{G} - \frac{1}{2}G + \frac{2}{3}B_{8_c8_c}$, 
where it has been assumed that the combination does not (significantly) run within the 
region of energy scale under consideration. 
By taking the measured $M({D_{s0}^+(2317)})_{\rm exp} = 2317.8\pm 0.6$ MeV~\cite{PDG16} 
as the input data, it can be determined as 
%%%%%%%%%%%%%%%%%%%%%%%%%%%%%%%%%%%%%%%%%%%%%%%%%%%%%%%%%%%%%%%%%%%%%%%%%%
\begin{equation}
2\hat{S} + \hat{G} - \frac{1}{2}G + \frac{2}{3}B_{8_c8_c} 
= (597.7\pm 30.5)\,\,{\rm MeV}.                           \label{eq:det-unknown-parameter}
\end{equation}
%%%%%%%%%%%%%%%%%%%%%%%%%%%%%%%%%%%%%%%%%%%%%%%%%%%%%%%%%%%%%%%%%%%%%%%%%%
Inserting the above result into Eq.~(\ref{eq:mass-formulae}), we can obtain mass values of 
open- and hidden-charm tetra-quark scalars as listed in Table I. 
As seen in the table, the present results on masses of open-charm sector  are not so  
much different from the previous ones in \cite{Terasaki-D_s}, while newly estimated 
masses of hidden-charm mesons are considerably lower than our earlier ones in 
\cite{KT-delta(3.2)}. 
%%%%%%%%%%%%%%%%%%%%%%%%%%%%%%%%%%%%%%%%%%%%%%%%%%%%%%%%%%%%%%%%%%%%%%%%%%
\begin{center}
\begin{table}[b]
\begin{quote}
%\caption{%
Table I. 
Estimated masses of open- and hidden-charm tetra-quark scalar mesons, where 
$\hat{D}^s$ and $\hat{\sigma}^{cs}$ are not included because each of them contains an 
$\{s\bar{s}\}$ pair. 
The measured $M({D_{s0}^+(2317)})_{\rm exp} = 2317.8\pm 0.6$ MeV~\cite{PDG16} and 
the running quark mass values which are listed in the text are taken as the input data. 
The listed errors are mainly from $m_c(\mu = m_c)$ in \cite{PDG16}. 
\vspace{2mm}\\
\end{quote}
%%%%%%%%%%%%%%%%%%%%%%%%%%%%%%%%%%%%%%%%%%%%%%%%%%%%%%%%%%%%%%%%%%%%%%%%%%
\begin{tabular}{|c|c|c|c|c|c|c|}
\hline
\begin{tabular}{c}
\\
$S$\\
\end{tabular} & 
\begin{tabular}{c}
\\
$I = 1$\\
\end{tabular} & 
\begin{tabular}{c}
\\
$I = 1/2$
\end{tabular} & 
\begin{tabular}{c}
\\
$I = 0$ 
\end{tabular}
& %\multicolumn{2}{|c|}{
\begin{tabular}{c}
Predicted mass \\
(MeV)%}
\end{tabular} 
& %\begin{tabular}{c}\vspace{-0mm}\\
\begin{tabular}{c}
Lowest OZI \\
allowed mode 
%\end{tabular} \vspace{-0mm}
\end{tabular}
& \begin{tabular}{c}
%\begin{tabular}{c}\vspace{-0mm}\\
Threshold \\ 
(MeV)
%\end{tabular}\vspace{-10mm}
\end{tabular}
%\\
%\cline{5-6}
%& & & & Present & Earier & & 
\\
\hline
\begin{tabular}{c}
\vspace{-3mm}\\
\vspace{-3mm} \\
$1$ \vspace{-7mm}\\
$$\vspace{-0mm}\\
\end{tabular} 
& \begin{tabular}{c}
\vspace{-4.5mm}\\
$\hat{F}_I$ 
\end{tabular}
&  & & \hspace{4mm}
$2317.8\pm \hspace{1mm}0.6\,\,(\ddagger)$ %& \hspace{8.5mm}$2320\,\,(\dagger)$ 
& $D_s^+\pi$ 
& $2106.5$
\vspace{-0.5mm}\\
\cline{2-7}
& & &
\begin{tabular}{c}
\vspace{-4mm} \\
$\hat{F}_0^+$
\end{tabular}  & \hspace{0mm}$2317.8 \pm \hspace{1mm}0.6$ %& \hspace{3.5mm}$2320$ 
&
\begin{tabular}{c}
\vspace{-4mm} \\
$D_s^+\eta$ %($D_s^{*+}\gamma$) 
\end{tabular} 
& $2516.1$ %$2112.3$ 
\\
\hline
$0$ & 
& \begin{tabular}{c}
\vspace{-4mm} \\
$\hat{D}$ 
\end{tabular}
& & $2213.5\pm 43.4$ %& \hspace{3.5mm}$2220$ 
& $D\pi$ 
& $2005.2$ 
\vspace{-0.5mm}\\
\hline
$-1$ & & & 
\begin{tabular}{c}
\vspace{-4mm} \\
$\hat{E}^0$
\end{tabular} 
& $2326.9\pm 43.1$ %& \hspace{3.5mm}$2320$ 
& $D\bar{K}$ 
& $2362.9$  
\vspace{-0.5mm}\\
\hline
$1$ 
&  & $\hat{\kappa}^c$ &  & $3079.2\pm 30.5$ %& $\sim 3400$ 
& $\eta_cK$ 
& \,$3479.4$
\vspace{-0mm}\\
\hline
\begin{tabular}{c}
\vspace{-4mm} \\
$0$ \vspace{-7mm}\\
$$\vspace{-2mm}\\
\end{tabular} 
& \begin{tabular}{c}
\vspace{-4mm}\\
$\hat{\delta}^c$ 
\end{tabular}
&  &  & $2980.6\pm 31.4$ %& $\sim 3300$ 
& $\eta_c\pi$ 
& $3121.4$
\vspace{-0.5mm}\\
\cline{2-7}
& & $$ & $\hat{\sigma}^{c}$ & $2980.6\pm 31.4$ %& $\sim 3300$ 
& $\eta_c\eta$ 
&\hspace{-1mm}$3531.3$
\\
\hline
\end{tabular} \vspace{1mm}
\\
($\ddagger$): Input data
\end{table}
\end{center}
%%%%%%%%%%%%%%%%%%%%%%%%%%%%%%%%%%%%%%%%%%%%%%%%%%%%%%%%%%%%%%%%%%%%%%%%%%

The iso-triplet $\hat{F}_I$ can decay dominantly into $D_s^+\pi$ through isospin 
conserving strong interactions, while the iso-singlet $\hat{F}_0^+$ should decay 
dominantly into $D_s^{*+}\gamma$, because of the hierarchy of hadron interactions, as 
discussed before. 
Therefore, the present approach expects observations of peaks around 2317 MeV not 
only in the $D_s^+\pi^0$ channel but also in the $D_s^{*+}\gamma$ (arising from 
$\hat{F}_I^+$ and $\hat{F}_0^+$, respectively), because of $M(\hat{F}_I) = M(\hat{F}_0)$.
Nevertheless, experiments observed no signal of the $D_s^{*+}\gamma$ peak. 
Even if $D_{s0}^+(2317)$ were identified to the ordinary $D_{s0}^{*+}$ as the 
$^3P_0\,\,\{c\bar{s}\}_{1_c}$, it also should decay dominantly into $D_s^{*+}\gamma$ 
for the same reason as the above~\cite{HT-isospin}.  
No signal of $D_{s0}^+(2317)$ in the $D_s^{*+}\gamma$ channel and no indicasion 
of  $D_{s0}^0(2317)$ and  $D_{s0}^{++}(2317)$ in $B$ decays are problems in our 
tetra-quark interpretations of $D_{s0}^+(2317)$, as discussed before. 

In regard to the non-strange 
%%%%%%%%%%%%%%%%%%%%%%%%%%%%%%%%%%%%%%%%%%%%%%%%%%%%%%%%%%%%%%%%%%%%%%%%%%
$\hat{D}\sim \{[cn]_{\bar{3}_c}^{1_s}[\bar{u}\bar{d}]_{3_c}^{1_s}\}_{1_c}^{1_s}$, 
%%%%%%%%%%%%%%%%%%%%%%%%%%%%%%%%%%%%%%%%%%%%%%%%%%%%%%%%%%%%%%%%%%%%%%%%%%
it can decay into $D\pi$ final states through isospin coserving strong interactions. 
However, it will be narrow for the same reason as the narrow width of 
$D_{s0}^+(2317)$~\cite{HT-isospin}. 
Therefore, the observed broad $D_0^*(2400)$ with a mass 
$M({D_0^*})_{\rm exp} = 2318\pm 29$ MeV~\cite{PDG16} will be the ordinary 
$^3P_0\,\,\{c\bar{n}\}_{1_c}$ meson. 
This implies that the tetra-quark $\hat{D}$ with the predicted mass 
$M({\hat{D}}) = 2213.5\pm 43.4$ MeV in Table I will be observed as a tiny peak on the 
lower tail of the observed broad $D\pi$ enhancement, $D_0^*$, because production rate 
of $\hat{D}$ will be much lower than that of $D_0^*$. 

The predicted mass of the truly exotic 
$\hat{E}^0 \sim \{[cs]_{\bar{3}_c}^{1_s}[\bar{u}\bar{d}]_{3_c}^{1_s}\}_{1_c}^{1_s}$ is lower than 
the $D\bar{K}$ threshold, (at most, a little bit higher than the threshold, even if the upper 
bound of the estimated large errors is taken), in the present approach. 
Therefore, its strong decay would be forbidden (or kinematically suppressed), and, in 
addition, its radiative decay is not allowed. 
This suggests that its search in inclusive $e^+e^-$ annihilation might be not very easy.   
If its strong decay is strictly forbidden, it might be detected, for example, in the 
successive weak decays, 
%%%%%%%%%%%%%%%%%%%%%%%%%%%%%%%%%%%%%%%%%%%%%%%%%%%%%%%%%%%%%%%%%%%%%%%%%%
$\bar{B}\rightarrow \bar{D}\hat{E}^0\rightarrow \bar{D}(\bar{K}\bar{K}\pi)^0$~\cite{Hyodo} 
and 
$\bar{B}\rightarrow \bar{D}\hat{E}^0\rightarrow \bar{D}(\bar{K}\bar{K}\bar{\ell}\nu_\ell)^0$, 
%%%%%%%%%%%%%%%%%%%%%%%%%%%%%%%%%%%%%%%%%%%%%%%%%%%%%%%%%%%%%%%%%%%%%%%%%%
where $\ell = \mu$ or $e$. 

The hidden-charm non-strange $\hat{\delta}^c$ and $\hat{\sigma}^c$ have a degenerate 
mass $M({\hat{\delta}^c}) = M({\hat{\sigma}^c})\simeq 3.0$ GeV, which is lower than the 
$\eta_c\pi$ threshold, in the present approach and therefore, their searches are expected 
to be done in OZI-suppressed channels and radiative ones. 
Here it should be noted that the above result is much lower than those from the other 
approaches, for example, $3723$ MeV as the 
mass of the lowest hidden-charm non-strange scalar meson from the diquark-antidiquark 
model~\cite{Maiani} and around 3700 MeV from a unitarized chiral model~\cite{Oset}. 
Therefore, confirmation of existence of $\hat{\delta}^c(3200)$ and $\hat{\sigma}^c(3200)$ 
will be useful to select a realistic model of multi-quark mesons. 

In summary the present results on masses of open-charm tetra-quark scalar mesons are 
not very much different from our previous estimates by using a naive quark counting, 
while those of hidden-charm sector are considerably lower than the earlier ones, so that 
expected decay property of the latter is now drastically changed. 
In particular, the predicted mass of $\hat{\delta}^{c}$ in the present work is much lower 
than our previous result, so that it now cannot decay into $\eta_c\pi$, in contrast to our 
previous work. 
As the result, its dominant decay will be the OZI-suppressed 
$\hat{\delta}^{c}\rightarrow \eta\pi$. 
This should be compared with the existing result from two photon collisions that a tiny 
$\eta\pi$ peak around 3.2 GeV was indicated but no signal in the $\eta_c\pi$ channel. 
(If its mass is truly $3.2$ GeV or higher, the experiment should have observed a peak 
at the same energy in the $\eta_c\pi$ channel.)
In addition, to search for hidden-charm tetra-quark scalar mesons, their OZI-suppressed 
and radiative decay channels would be important, because their OZI-allowed hadronic 
decays are kinematically suppressed or not allowed. 
However, detailed studies of these decays are left intact as our future subjects. 

With regard to the charm-strange scalar $D_{s0}^+(2317)$, no indication of  
$D_{s0}^+(2317)\rightarrow D_s^{*+}\gamma$ and no signal of its neutral and doubly 
charged partners in $B$ decays would be a serious dilemma in its tetra-quark 
interpretations, unless production rates for its neutral and doubly charged partners in 
addition to its iso-singlet one are unexpectedly suppressed. 
In contrast, there exist arguments that hadronic loop contributions might induce a strong 
suppression of the $D_s^{*+}\gamma$ decay and/or an extraordinary enhancement of 
the isospin nonconserving $D_s^+\pi^0$ decay of iso-singlet $DK$ 
molecule~\cite{Faessler,Hanhart}. 
Therefore, one might consider that the $DK$ molecular picture of $D_{s0}^+(2317)$ is 
favored by experiments.
%the above experimental results that $D_{s0}^+(2317)$ is observed in the 
%$D_s^+\pi^0$ channel, while no signal in the radiative $D_s^{*+}\gamma$. 
Nevertheless, these analyses contain various adjustable parameters and expediential  
tunings of their values are needed to solve the above puzzle. 
In \cite{Faessler}, for example, a size parameter $\Lambda_{D_{s0}(2317)}$ describing the  
size of $D_{s0}^+(2317)$ has been introduced, and it has been discussed that the above 
puzzle can be solved for $\Lambda_{D_{s0}(2317)} > 1$ GeV. 
However, this condition implies that the size of  %$\Lambda_{D_{s0}(2317)}^{-1} < 0.2$ fm 
the molecular $DK\sim D_{s0}^+(2317)$ is more compact than the constituent $K$ meson 
with its charge radius $\sqrt{\langle r\rangle^2} \simeq 0.58$ fm~\cite{Amendolia}, i.e., 
the constituent $K$, which is treated as a point particle in the analysis, has a size larger 
than the molecular $D_{s0}^+(2317)$. 
On the other hand, in a more recent analysis~\cite{Hanhart} in which more restricted loop 
contributions have been taken into account, it has been argued that the isospin 
nonconsrving $D_s^+\pi^0$ decay of $D_{s0}^+(2317)$ is sensistive to tuning of adjustable 
parameters involved, and the resulting ratio of rates 
%%%%%%%%%%%%%%%%%%%%%%%%%%%%%%%%%%%%%%%%%%%%%%%%%%%%%%%%%%%%%%%%%%%%%%%%%%
$R = \Gamma(D_{s0}^+(2317)\rightarrow D_s^+\gamma)
       /\Gamma(D_{s0}^+(2317)\rightarrow D_s^+\pi^0)$ 
%%%%%%%%%%%%%%%%%%%%%%%%%%%%%%%%%%%%%%%%%%%%%%%%%%%%%%%%%%%%%%%%%%%%%%%%%%
with a makeshift tuning of their values can be considerably lower than unity, while it 
cannot completely satisfy its measured restriction $R_{\rm exp} < 0.059$~\cite{PDG16}. 
In this case, it might not be easy to understand why no indication of $D_{s0}^+(2317)$ has  
been observed in the $D_s^{*+}\gamma$ channel. 
Thus, it seems to be still unclear at the present stage if the iso-singlet $DK$ molecular 
picture of $D_{s0}^+(2317)$ is acceptable. 

As seen above, physics of $D_{s0}^+(2317)$ and its partners seems to be still confusing. 
To determine their structure, more experimental and theoretical investigations will be 
needed. \vspace{-0mm}

%%%%%%%%%%%%%%%%%%%
\newpage
%%%%%%%%%%%%%%%%%%%%%%%%%%%%%%%%%%%%%%%%%%%%%%%%%%%%%%%%%%%%%%%%%%%%%%%%%%
\section*{Acknowledgments} \vspace{-0mm}
{The author would like to thank Prof.~T.~Hyodo, YITP (Yukawa Institute for Theoretical 
Physics, Kyoto University) for valuable discussions and comments. 
He also would like to appreciate Prof.~H.~Kunitomo, YITP for careful reading of the 
manuscript.} 
%%%%%%%%%%%%%%%%%%%%%%%%%%%%%%%%%%%%%%%%%%%%%%%%%%%%%%%%%%%%%%%%%%%%%%%%%%

%%%%%%%%%%%%%%%%%%%%%%%%%%%%%%%%%

%\end{references}
%%%%%%%%%%%%%%%%%%%%%%%

\begin{thebibliography}{99}
\setlength{\itemsep}{2pt} 

{ %\normalsize 


\bibitem{D_{s0}(2317)-exp}
%\bibitem{Babar-D_s}
B.~Aubert et al., the Babar Collaboration, Phys. Rev. Lett. {\bf 90}, 
242001 (2003); 
%\bibitem{CLEO-D_s}
D.~Besson et al., the CLEO Collaboration, Phys. Rev. D {\bf 68}, 032002 (2003); 
%hep-ex/0305017; 
%\bibitem{Belle-D_s} 
P.~Krokovny et al., the Belle Collaboration, Phys. Rev. Lett. {\bf 91}, 262002  (2003). 
%; hep-ex/0307041. 

\bibitem{Swanson} 
Multi-quark interpretations of new resonances %including meson-meson molecules 
are summarized, for example, in E.~S.~Swanson, Phys. Rep. {\bf 429}, 243 (2006).  
See also references quoted therein.   

\bibitem{Cheng-Hou} 
H.-Y.~Cheng and W.-S.~Hou, Phys. Lett. {\bf B566}, 193 (2003). 
%; hep-ph/0305012.

\bibitem{Terasaki-D_s} 
K.~Terasaki, Phys. Rev. D {\bf 68}, 011501(R) (2003). 

\bibitem{Uehara-delta(3.2)}
S.~Uehara et al., the Belle Collaboration, Phys. Rev. D {\bf 80}, 032001 (2009). 

\bibitem{KT-delta(3.2)}
K.~Terasaki, Prog. Theor. Phys. {\bf 121}, 211 (2009); arXiv:0805.4460 [hep-ph]. 
%; K.~Terasaki, arXiv:1304.7080. 

\bibitem{Uehara-eta_c-pi} 
S.~Uehara, private communications. 

\bibitem{OZI} 
S.~Okubo, Phys. Lett. {\bf 5},165 (1963); G.~Zweig, CERN Report No. TH401 (1964); 
J.~Iizuka, K.~Okada and O.~Shito, Prog. Theor. Phys. {\bf 35},1061 (1965). 

\bibitem{Jaffe} R.~L.~Jaffe, Phys. Rev. D {\bf 15}, 267 and 281(1977). 

\bibitem{Hori} 
M.~Y.~Han and Y.~Nambu, Phys. Rev. {\bf 139}, B1006 (1965); 
S.~Hori, Prog. Theor. Phys. {\bf 36}, 131 (1966). 

\bibitem{TM}
K.~Terasaki and B.~H.~J.~McKellar, Prog. Theor. Phys. {\bf 114}, 205 (2005); 
hep-ph/0501188. 

\bibitem{PDG96}
R.~M.~Barnet et al., the Particle Data Group, Phys. Rev. D {\bf 54}, 1 (1996). 

\bibitem{Leutwyler} 
H.~Leutwyler and M.~Roos, Z. Phys. C {\bf 25}, 91 (1984).

\bibitem{FNAL-E687}
M.~S.~Nehring, the Fermilab E687, Nucl. Phys. B (Proc.Suppl.) {\bf 55A}, 131 (1997). 

\bibitem{CLEO-FF}
J.~Barnett et al., the CLEO, Phys. Lett. {\bf B405}, 373 (1997).  

\bibitem{PDG06} W.-M.~Yao et al., the Particle Data Group, J. Phys. G {\bf 33}, 1 (2006), 
and references quoted therein. 

\bibitem{E791} 
E.~M.~Aitala et al., the Fermilab E791 Collaboration, Phys. Rev. Lett. {\bf 89}, 121801 
(2002). 

\bibitem{HT-isospin} 
A.~Hayashigaki and K.~Terasaki, Prog. Theor. Phys. {\bf 114}, 1191 (2006); 
hep-ph/0410393. 

\bibitem{Dalitz}
R.~H.~Dalitz and F.~Von Hippel, Phys. Lett. {\bf 10} (1964), 153. 

\bibitem{search-for-double-charge} %No signal of $D_s^{++}(2317)$ and $D_s^0(2317)$
S.-K.~Choi et al., the Belle Collaboration, Phys. Rev. D {\bf 91} 9, 092011(2015) ; 
Phys. Rev. D {\bf 92} 3, 039905 (2015). %; arXiv:1504.02637. 

\bibitem{KT-production}
K.~Terasaki,  Prog. Theor. Phys. {\bf 116}, 435 (2006); hep-ph/0604207; 
%. ; arXiv:1604.06161 [hep-ph]
%
%\bibitem{KT-dilemma}K.~Terasaki, 
arXiv:1604.06161 (unpublished). 

\bibitem{KNR} 
M.~Karliner, S.~Nussinov and J.~L.~Rosner, Phys. Rev. D {\bf 95}, 034011 (2017). 
%; arXiv:1611.00348. 

\bibitem{PDG16}
C.~Patrignani et al., the Particle Data Group, Chin. Phys. C {\bf 40}, 100001 (2016). 

\bibitem{m_c(3GeV)}
K.~G.~Chetyrkin, J.~H.~Kuhn, A.~Maier, P.~Maierhofer, P.~Marquard, M.~Steinhauser, 
C.~Sturm, Phys. Rev. D {\bf 96}, 116007 (2017). %;  arViv:1710.04249 [hep-ph]. 

\bibitem{KT-decomp}
K.~Terasaki, hep-ph/0512285 (unpublished); Eur. Phys. J. {\bf A31}, 676 (2007); 
hep-ph/0609233. 
%;  Prog. Theor. Phys. {\bf 121}, 211 (2009); arXiv:0805.4460 [hep-ph].

\bibitem{Hyodo}
T.~Hyodo, private communications. 

\bibitem{Maiani}
L.~Maiani, F.~Piccinini, A.~D.~Polosa and V.~Riquer, Phys. Rev. D {\bf 71}, 014028 (2005). 

\bibitem{Oset}
D.~Gamermann, L.~R.~Dai and E.~Oset, Phys. Rev. C {\bf 76}, 055205 (2007). 

\bibitem{Faessler}
A.~Faessler, T.~Gusche, V.~E.~Lyuboviskij and Y.-L.~Ma, Phys. Rev. D {\bf 76}, 014005 
(2007). 

\bibitem{Hanhart} 
M.~Cleven, H.~W.~Griesshammer, F.-K.~Guo, C.~Hanhart, U.-G.~Meissner, 
Eur. Phys. J. {\bf A50}, 149 (2014). 

\bibitem{Amendolia}
S.~R.~Amendolia et al., Phys. Lett. {\bf B178}, 435 (1986). 

%\bibitem{HT-QCDSR}
%A.~Hayashigaki and K.~Terasaki, hep-ph/0411285 (unpublished). 

%\bibitem{CLEO-csbar-scalar}
%Y.~Kubota et al., the CLEO collaboration, Phys. Rev. Lett. {\bf 72}, 1972 (1994). 

%\bibitem{Olsen-a_0-f_0-mixing} 
%S.~L.~Olsen, arXiv:1203.4297[nucl-ex] and references quoted therein. 

%\bibitem{VMD} 
%M.~Gell-Mann and F.~Zachariasen, Phys. Rev. {\bf 124}, 953 (1961); 
%J.~J.~Sakurai, {\it Currents and Mesons} (Chicago, Ill., 1969). 

%\bibitem{KT-hidden-charm-scalar}
%K.~Terasaki, Prog. Theor. Phys. {\bf 121}, 211 (2009); arXiv:0805.4460 [hep-ph].

%\bibitem{Terasaki-VMD} 
%K.~Terasaki, Lett. Nuovo Cimento {\bf 31}, 457 (1981); 
%Il Nuovo Cimento {\bf 66A}, 475 (1981). The experimental data used in 
%these articles are now updated and the phase convention of $\phi$ 
%and $\psi$ is changed in this note.  

%\bibitem{Belle-X-rho} 
%S.-K.~Choi et al., Belle Collaboration, Phys. Rev. Lett. {\bf 93}, 
%26200 (2003). 

%\bibitem{Terasaki-X} 
%K.~Terasaki, Prog. Theor. Phys. {\bf 118}, 821 (2007); 
%hep-ph/0706.3944.

%%%%%%%%%%%%%%%%%%%%%%%%%%%%%%%%%%%%%%%%%%%%%%%%%%%%%%%%%%%%%%%%%%%%%%%
% Succeeding analyses
%\bibitem{Achasov} 
%N.~N.~Achasov and V.~N.~Ivanchenko, Nucl. Phys. B {\bf 315} (1989), 
%465;
%N.~N.~Achasov and A.~V.~Kiselev, hep-ph/0512047 and references therein. 
%\bibitem{Bugg} 
%D.~V.~Bugg, hep-ex/0510014 and references therein. 
%\bibitem{Maiani} 
%L.~Maiani, F.~Piccinini, A.~D.~Polosa and V.~Riquer, 
%Phys. Rev. Lett. {\bf 93} (2004), 212002.
%\bibitem{CT} 
%F.~E.~Close and N.~A.~T\"ornquvist, J. Phys. G {\bf 28}, R249 (2002). 
%\bibitem{Suganuma} 
%H.~Suganuma, K.~Tsumura, N.~Ishii and F.~Okiharu, hep-lat/0509121. 
%%%%%%%%%%%%%%%%%%%%%%%%%%%%%%%%%%%%%%%%%%%%%%%%%%%%%%%%%%%%%%%%%%%%%%%

%\bibitem{ECT-talk}  
%K.~Terasaki, hep-ph/0512285. 

%\bibitem{Lisbon}
%K.~Terasaki, hep-ph/0804.2295 (talk given at SCADRON70, the workshop 
%on ``Scalar Mesons and Related Topics'', 11 -- 16 February 2008, 
%IST-Lisbon, Lisbon, Portugal). In this article, the production mechanism of 
%charm-strange tetra-quark mesons in Ref.~\cite{production} has been 
%revised. 

%\bibitem{Babar-search} 
%B.~Aubert et al., the Babar Collaboration, hep-ex/0604030. 

%\bibitem{production} 
%K.~Terasaki, Prog. Theor. Phys. {\bf 116}, 435 (2006);  
%hep-ph/0604207;\\ 
%Eur. Phys. J. A {\bf 31}, 676 (2007); hep-ph/0704.3299.  

%\bibitem{Belle-X-omega} 
%K.~Abe et al., Belle Collaboration, hep-ex/0505037. 

%\bibitem{Belle-X-J^P} 
%K.~Abe et al., Belle Collaboration, hep-ex/0505038. 

%\bibitem{hard-pion} 
%K.~Terasaki, S.~Oneda and T.~Tanuma, Phys. Rev. D {\bf 29}, 456 (1984). 

%\bibitem{suppl} S.~Oneda and K.~Terasaki, Prog. Theor. Phys. Suppl. 
%No. {82}, 1 (1985) and references quoted therein. 

%\bibitem{TM} 
%K.~Terasaki and Bruce H J McKellar, Prog. Theor. Phys. {\bf 114}, 205 
%(2005).
}
%%%%%%%%%%%%%%%%%%%%%%%%%%%%%%%%%%%%%
\end{thebibliography}
\end{document}